\documentclass[aps,pre,showkeys,showpacs,floatfix, twocolumn]{revtex4} 
\usepackage{graphicx}
\usepackage[ansinew]{inputenc}
\usepackage[tbtags]{amsmath}
\usepackage{amssymb}
\usepackage{float}

\newcommand{\beq}{\begin{equation}}
\newcommand{\eeq}{\end{equation}}
\newcommand{\bc}{\begin{center}}
\newcommand{\ec}{\end{center}}
\newcommand{\eps}{\epsilon}
\newcommand{\bx}{\mathbf{x}}
\newcommand{\by}{\mathbf{y}}
\newcommand{\bz}{\mathbf{z}}

\begin{document}
\title{Optimal Markov Approximations and Generalized Embeddings}
\author{Detlef Holstein}
\email{holstein@pks.mpg.de}
\author{Holger Kantz}
\affiliation{Max Planck Institute for the Physics of Complex Systems,
N\"othnitzer Str.\ 38, 01187 Dresden, Germany}
\date{\today}

\begin{abstract}
Based on information theory, we present a method to determine 
an optimal Markov approximation for modelling and prediction
from time series data.
The method finds a balance between minimal modelling
errors by taking as much as possible memory into account and
minimal statistical errors by working in embedding spaces of 
rather small dimension. 
A key ingredient is an estimate of the statistical error of entropy 
estimates. The method is illustrated with several examples and 
the consequences for prediction are evaluated by means of 
the root mean squard prediction error for point prediction.
\end{abstract} 


\maketitle

\section{Introduction}
Given is a
univariate time series $\{x_i: i=1,\ldots,N\}$, obtained from the
time evolution of some deterministic or stochastic dynamical system by
applying a scalar measurement function to the state vectors of this
system. We will assume that the measurements are equidistant in time. 
A meanwhile standard approach to the modelling and prediction
based on univariate time series data starts from the construction of a
multidimensional state space. Commonly used is the time delay embedding
space.
In the case of deterministic dynamics, the Takens theorem \cite{takens81} 
states that if the embedding dimension $m$ satisfies $m>2 D_f$, where
$D_f$ is the fractal dimension of the attractor, then m-dimensional 
delay vectors $(x_{n-k(m-1)}, x_{n-k(m-2)}, \ldots, x_n)$ with delay $k$
can be uniquely mapped onto the non-observed state vectors. 
Hence the process in the m-dimensional delay embedding space is
deterministic in the sense of the existence and uniqueness of the
solution of the initial value problem. In special cases, smaller values 
of $m$ might be sufficient for reconstruction of the underlying dynamics. 

As it has been argued recently \cite{paparella97, kantzholst04}, 
also the modelling and prediction of time series data
from stochastic processes can profit from the concept of state space
reconstruction: In an ideal situation, there exists a time delay
embedding space, in which the stochastic dynamics is Markovian
of (possibly higher) order $m$, i.e.,
in which the conditional probability density function (pdf)
$\rho(x_n|x_{n-m}, \ldots, x_{n-1})$ to find a given future
value cannot be made narrower by including more past values 
into the condition. In the framework of time series analysis, 
the conditional pdf has to be estimated from the data. 
This can be done by either estimating conditional probabilities 
through binning and counting \cite{kantzholst04},
or by kernel estimators \cite{silverman92}. 
Two consequences arise: These estimates are
subject to statistical errors, and a length scale $\eps$ is introduced, 
i.e., the estimated conditional probabilities do not vary as a function 
of the condition $(x_{n-m}, \ldots, x_{n-1})$ on length scales smaller 
than $\eps$. The statistical error is a function of not only the 
dataset size $N$ but also of the spatial resolution $\eps$. 
When models have to be fitted to observed data, model parameters 
are to be determined. The estimated conditional probabilities can
be interpreted as the model parameters of a Markov model. 
In data analysis tasks the Markov order $m$, however, usually is not 
a priori known, and has to be obtained from the data.

In both the deterministic and the stochastic cases, finding the 
suitable embedding dimension is one of the practical issues.
In the stochastic case the embedding dimension can be associated 
with the number of time steps of nonvanishing condition, which 
under absence of intermediate time steps of vanishing condition
reduces to the Markov order $m$. Whereas for the deterministic case
mathematically rigorous results \cite{saueryorkecasdagli91}
as well as numerically efficient and reliable algorithms
\cite{grassproc83a} exist, for the stochastic case only
statistically demanding tests of the Chapman Kolmogorov equation are
currently in use \cite{friedrich97}. 
In both cases, there exists the practical problem
that from a theoretical point of view the embedding dimension
for the process could be very large. If the amount of data is
insufficient in view of statistical robustness of either the
algorithms to determine empirically the embedding dimension or of
estimates of, e.g., model parameters in 
a corresponding space, then a
high dimensional model is practically irrelevant, even if
theoretically justified. Hence, in many situations an effective model
and a different embedding dimension might be superior to the 
model advised by the structure of the underlying dynamics.
We will illustrate this
statement later and we will convince the reader of its relevance. 

When identifying the optimal embeddings, i.e., when looking for
the optimal Markov approximations, 
we have to take into account two types of errors. 
The first one is a modelling error, which we make if we ignore 
components in the past of the time series which are relevant for its
future. In the deterministic setting, this would mean that we use an
embedding dimension which is too small. In the stochastic setting, it
means that the Makovianity of in general higher order 
or the cardinality of time steps of nonvanishing condition
is not fully captured by the chosen embedding space. 
The second error is a statistical error. Regardless of which quantity
is estimated from a finite dataset, its value is always subject to a
statistical error. In the context of prediction and modeling, the
corresponding samples are usually obtained from 
neighbourhoods of delay vectors.
Sample sizes are small and correspondingly statistical errors large,
if we work in embedding spaces whose dimension is too large compared
to the amount of available data and compared to the diameter of
neighbourhoods, i.e., locality of the estimate. 

Hence, what we are proposing here with the intention of a rather general 
applicability is a concept to identify optimal resolution-dependent 
Markov approximations, in which the combined effect of modeling errors 
and statistical errors is minimal.

Practically, we will relate the modelling error to the discrepancy of 
conditional entropies from entropies with sufficient conditioning.
The statistical error of a model will be related to the statistical 
error in entropy estimation. Therefore, we will carry out explicitly
error estimates for entropy estimations. On the route of searching for
models which capture the memory of a process but involve an as small
as possible embedding dimension, we will consider also non-standard 
so-called perforated embeddings, namely those, where the temporal 
spacings between successive elements of a delay vector are not 
identical for all pairs of adjacent components. Such embeddings 
were also discussed in \cite{garcia05} and \cite{pecora06}.
This paper makes a new suggestion how to find optimal ones.

In the next section the basic quantities of information theory 
are introduced, which in the development of the criteria for 
optimal Markov approximations play a certain role.
In Sec.\ref{sec:entropestimandstatisterrors} we remind a widely used 
procedure for the estimation of entropies and discuss the statistical errors 
in numerical estimation of the correlation entropy. 
In Sec.\ref{sec:optimusualmarkovapprox} a novel method for 
the selection of usual Markov approximations is presented, 
but it is immediately 
pointed out that the framework has to be generalized in order to be 
suitable for arbitrary dynamics.
A unified notation for entropies in the time series analytical framework
suitable for the treatment of variable future lead times, 
jointly conditioned joint entropies, noncausal conditionings, 
downsampling and arbitrary omissions in conditionings
is introduced in Sec.\ref{sec:perforatedness},
which remedies the formerly mentioned problems.
The notion of perforatedness is introduced.
In Sec.\ref{sec:optgeneralizmarkovapprox} we present the
method to identify optimal generalized Markov approximations
as a function of the data accuracy $\eps$
for a given time series of fixed length $N$. 
Subsequently, the success of the introduced criterion for the 
determination of optimal perforated Markov approximations is 
illustrated for several model processes with memory in 
Sec.\ref{sec:examplesperfmarkovapprox}.
We show that indeed the
theoretically optimal embedding of the process 
from the dynamical law behind the generated data sets
is not necessarily the optimal state space representation 
of a finite time series for all resolutions.
Furthermore the dependence on the length of the underlying
dataset is discussed in detail. 
Some consequences for prediction with the example 
of the generalized Henon map
are outlined in Sec.\ref{sec:conseqforpred}.
In Sec.\ref{sec:conclusion} the results of this paper are concluded.

\section{Relevant quantities of information theory}
The resolution($\eps$)-dependent joint Renyi block entropy of order $q$
is given by
\beq
\label{eq:renyiblockentropy}
H_m^{(q)}(\eps)=\frac{1}{1-q}\ln \left[\sum_{l_1=1}^{L_1(\eps)} \ldots
\sum_{l_m=1}^{L_m(\eps)} {(p_{l_1, \ldots ,l_m}(\eps))}^q \right]\; .
\eeq
It estimates the joint uncertainty of random variables corresponding 
to $m$ successive time steps of a time series. In case of dependences 
of random variables a conditional probability distribution is narrower 
than the corresponding unconditioned probability distribution. Further 
conditioning into the past in general further decreases the width of 
the distribution and the uncertainty of the outcome of the random 
experiment. This behaviour can also be quantified with 
{\emph{conditional entropies}} defined by
\beq
\label{eq:nonperfcondent}
H_{1|m}(\eps):=H_{m+1}(\eps)-H_m(\eps) \; ,
\eeq
as the difference of joint block entropies with different block length.
In this formula and the subsequent ones the Renyi order $q$ is 
notationally omitted. Conditional entropies are interpreted as the 
remaining uncertainty after having used the information from the 
chosen conditioning. In case of maximal, i.e. infinite conditioning, 
the unreducable uncertainty is obtained as
\beq
\label{eq:infiniteconditcondent}
H_{1|\infty}(\eps)\equiv \lim_{m \rightarrow \infty} H_{1|m}(\eps)\; .
\eeq
The {\emph{redundancy}} is defined by
\beq
\label{eq:redundancy}
R_{m}(\eps) := H_1(\eps) - H_{1|m}(\eps)\; ,
\eeq
and hence is interpreted as the uncertainty reduction of the immediate 
future random variable from conditioning on the adjacent $m$ past time steps.
The quantity
\beq
\label{eq:ignoredmemory}
Q_m(\eps):=H_{1|m}(\eps)-H_{1|\infty}(\eps)
\eeq
is called {\emph{ignored memory}}. 
It is the in principle accessible, but renounced 
uncertainty reduction of the immediate future random variable.
From combining Eq.(\ref{eq:redundancy}) and Eq.(\ref{eq:ignoredmemory})
it is possible to see that the total uncertainty of a single random 
variable is decomposable according to
\beq
H_1(\eps)=H_{1|\infty}(\eps)+R_m(\eps)+Q_m(\eps)\; .
\eeq

\section{Estimation of entropic quantities and 
statistical errors of correlation entropies}
\label{sec:entropestimandstatisterrors}
Our method for finding optimal Markov approximations 
will find a balance between maximal uncertainty reduction of
future values of the time series and minimized statistical errors. 
To this end, we will discuss here the estimation of entropies 
from finite time series and in particular the statistical errors 
involved in this estimation. Because it is the statistically
most robust and algorithmically most convenient quantity, we will
concentrate here on the order-2 Renyi entropy $H^{(q=2)}=-\ln\sum p_i^2$,
which is estimated from the Grassberger-Procaccia correlation sum
\cite{grassproc83a}
\beq
\label{eq:correlationsum}
C_m^{(2)}(N,\eps) = \frac{1}{N(N-1)} \sum_{i=1}^N \sum_{k\ne i}
\Theta(\eps - \|\bx_i - \bx_k\|)
\eeq
by
\beq
\label{eq:renyizweientropy}
H_m^{(2)}(N,\eps)=-\ln C_m^{(2)}(N, \eps)\; .
\eeq
We use the maximum norm in the argument of the Heaviside function $\Theta$.
The conditional entropy of Eq.(\ref{eq:nonperfcondent}), 
our construction element for ignored memory and redundancy,
is obtained from
\beq
\label{eq:condentropy}
H_{1|m}^{(2)}(N, \eps)= \ln C_m^{(2)}(N,\eps)
-\ln C_{m+1}^{(2)}(N,\eps)\; .
\eeq
As it was shown by Grassberger \cite{grassberger03}, the correlation
sum does not suffer from systematic finite sample effects, i.e., it is
an unbiased estimator of the correlation integral. Consequently, the
mean value of
estimated quantities such as the correlation entropy or the correlation
dimension on data sets of fixed size $N$ for arbitrarily small $\eps$
will be correct, as long as the combination $(N,\eps)$ is such
that the correlation sum is non-zero. However, each individual result
is subject to statistical errors. In the following we want to estimate 
the magnitude of these errors. 

To begin with, it is introduced the random variable 
$W_m(\eps, \bx_i)$ for the number of 
similar vectors $\bx_k$ of $\bx_i$ with distance 
smaller than $\eps$ according to the chosen norm, i.e., 
the random variable for the cardinality of the set \linebreak
$\{\bx_k \in {\mathcal{U}}(\eps, \bx_i): k\in\{1, \ldots , N-m\} \}$,
where the $\eps$-neighborhood of the vector $\bx$ is defined by
\beq
\label{eq:epsneighborhood}
{\mathcal{U}}(\eps, \bx) 
:=\{\bz: \|\bz-\bx\|< \eps\}\; .
\eeq
$W_m(\eps, \bx_i)$
is distributed according to a Binomial distribution.
For a given dataset, the realization of $W_m(\eps, \bx_i)$
is given by 
\beq
\label{eq:realizcardininneighborhood}
w_m(N, \eps, \bx_i):=
\sum_{k=1; k\neq i}^N\Theta(\eps - \|\bx_i-\bx_k\|) \; . 
\eeq
With this expression the correlation sum (Eq.(\ref{eq:correlationsum}))
can be written as
\beq
\label{eq:corrsumwithhelpquantity}
C_m^{(2)}(N, \eps)
= \frac{1}{N(N-1)}\sum_i w_m(N, \eps, \bx_i) \; .
\eeq
Except for very large $\eps$ or extremely small $N$ 
the distribution of the random variable $W_m(\eps, \bx_i)$
can be excellently approximated by a Poisson distribution. This leads to
the property
\beq
\operatorname{Var}(W_m(\eps, \bx_i))
= E(W_m(\eps, \bx_i)) \; ,
\eeq
and therefore
\beq
\label{eq:approxrelation}
\Delta w_m(N, \eps, \bx_i) \approx 
\sqrt{w_m(N, \eps, \bx_i)} \; .
\eeq
Assuming mutual independence of $W_m(\eps, \bx_i)$ and using 
the standard rules for error propagation (additivity of the variances)
as well as the approximate relation (\ref{eq:approxrelation}),
the statistical error of the correlation sum is estimated by
\begin{align}
\label{eq:C2error}
\Delta C_m^{(2)}(N, \eps) 
& = \frac{1}{N(N-1)} \sqrt{\sum_i 
(\Delta w_m(N, \eps, \bx_i))^2} \nonumber \\
& \approx \frac{1}{N(N-1)} \sqrt{\sum_i 
w_m(N, \eps, \bx_i)} \; .
\end{align}
Thus it can be
computed by using the non-normalized correlation sums, which are needed
anyway to estimate entropies.
From Eqs.(\ref{eq:renyizweientropy}), (\ref{eq:corrsumwithhelpquantity})
and (\ref{eq:C2error}) the statistical error of the 
Renyi entropy can be calculated as
\beq
\label{eq:errorjointentropyestim}
\Delta H_m^{(2)}(N, \eps)
 =\frac{\Delta C_m^{(2)}(N, \eps)}{C_m^{(2)}(N, \eps)} 
 \approx \frac{1}{\sqrt{\sum_i w_m(N, \eps, \bx_i)}}\; ,
\eeq
and the statistical error of the usual conditional entropy is obtained from
\begin{align}
\Delta H_{1|m}^{(2)}(N, \eps)
&=\sqrt{[\Delta H^{(2)}_{m+1}(N, \eps)]^2
+[\Delta H^{(2)}_m(N, \eps)]^2} \nonumber \\ 
&\approx \sqrt{
\frac{1}{\sum_i w_{m+1}(N, \eps, \by_i)}
+
\frac{1}{\sum_i w_m(N, \eps, \bx_i)} 
} \; .
\end{align}
Further error propagation for the estimated redundancy 
\begin{align}
\label{eq:statisterrofnonperfredund}
\Delta & R_m \! (N, \eps) \nonumber \\
& \! = \! \sqrt{[\Delta \! H_1 \! (N, \eps)]^2
\!\! + \! [\Delta \! H_{m+1} \! (N, \eps)]^2
\!\! + \! [\Delta \! H_m \! (N, \eps)]^2} 
\end{align}
is possible in the same way. This quantity will be needed for the criterion
given in Eq.(\ref{eq:critusualmarkovapprox}).

The assumptions entering the arguments for usual error propagation are: 
\begin{enumerate}
\item Independence of the random variables
\item Gaussian error statistics 
\item Errors are small so that nonlinear expressions can be
  approximated by first order Taylor expansions around the mean.
\end{enumerate}
The authors are aware of the fact that item 1 is violated, 
since if $\|\bx_i - \bx_{i'}\|< \eps$, then
the phase space points have overlapping
neighborhoods and $W_m(\eps, \bx_i)$ and $W_m(\eps, \bx_{i'})$ 
are not independent of each other. 
Correlations among the $W_m(\eps, \bx_i)$ yield a smaller 
effective sample size, such that Eq.(\ref{eq:errorjointentropyestim})
is an underestimation of the true statistical error of entropies.
The violation of the assumption of item 1, 
however, becomes the less relevant the smaller $\eps$, since then 
the overlap of neighborhoods decreases.
Item 2 is violated, since the error statistics of our basic random
variables $W_m(\eps, \bx_i)$ is explicitly non-Gaussian. This 
violation becomes the stronger the smaller the values of 
$w_m(N, \eps, \bx_i)$ become, i.e., for small $\eps$. 
Nevertheless, in spite of those arguments, usual error propagation 
is used as an approximation of the true errors of the estimation 
of entropic quantities.

Except for Eq.(\ref{eq:estimperfentrop}) 
in the following the dependence on the length $N$ of the dataset
will only be shown for the statistical errors, since 
for the expectation value of entropies and derived quantities 
there is no dependence on \nolinebreak $N$.

\section{A novel criterion for usual Markov approximations}
\label{sec:optimusualmarkovapprox}
As already mentioned in the introduction, there are two kinds of 
errors involved in our strategy for the determination 
of optimal Markov approximations:

First, there is a modelling error.
If a Markov approximation is carried out, typically information 
about the future is truncated,
which is not anymore available for uncertainty reduction.
This renounced uncertainty reduction can be quantified by 
the ignored memory $Q_m$ given in Eq.(\ref{eq:ignoredmemory}).
The value of $Q_m$ should be small. It is the smaller (or remains the same)
the more components in the past are taken into account, i.e., the 
higher the order $m$ of the Markov approximation.
Naively, one could be tempted to demand that the ignored memory in 
optimal Markov approximations should vanish, but in case of
infinite range of memory in the past the resulting 
Markov order would be infinite, what cannot be desired with respect
to practical applications.

Second, a statistical error has to be discussed.
There is an unavoidable statistical error in the estimation of entropies,
which is propagated to a nonvanishing statistical error 
of the performed uncertainty reduction from conditioning.
This statistical error quantified by $\Delta R_m$
given in Eq.(\ref{eq:statisterrofnonperfredund})
describes the unreproducibility of uncertainty reduction.
Also this term should be rather small in order to make the 
uncertainty reduction confident. $\Delta R_m$ increases with larger 
Markov order $m$, because less neighbors are found 
in the estimation of the correlation sum under the more restrictive conditions.
Demanding only the minimization of the statistical error 
of the redundancy in a criterion for optimal Markov approximations
would thus lead to empty conditionings.
It is intuitively clear that also this can in general not be 
a reasonable solution.

Since in contrast to $Q_m$ the term $\Delta R_m$ is the smaller the fewer 
past components are taken into account, the reduction of both errors are 
complementary demands and one faces an optimzation problem.
The aim is now to give a criterion such that for arbitrary dynamics 
a resolution-dependent optimal Markov approximation can be found.
The ad hoc choice for such a criterion reads:
\beq
\label{eq:critusualmarkovapprox}
m_{\operatorname{opt}}(\eps)=\max\{m \in {\mathbb{N}}_0 : \Delta R_m(N, \eps) < Q_m(\eps)\} \; .
\eeq
The reason for the criterion can be understood from the following:
The maximal memory, which in a senseful way to take into account
is restricted by the condition that the statistical error of redundancy, 
i.e., the statistical error of the uncertainty reduction has to be 
smaller than the ignored memory.
Otherwise the ignored memory is anyway not anymore resolvable
by enlargement of the order of the Markov approximation.
It is used that the statistical error of the redundancy increases 
with the Markov order, whereas the ignored memory decreases 
with the Markov order. 
Hence starting from the smallest possible Markov order $m$ 
it is increased as long as the statistical error of the redundancy
remains smaller than the ignored memory. 

The whole reasoning is resolution-dependent. In the Markov model
conditional probabilities
\begin{align}
\label{eq:condprobs}
& p_{l_n|l_{n-m}, \ldots, l_{n-1}}(\eps) \nonumber \\
& \quad =
\negthickspace 
\int\limits_{B_{l_{n-m}, \ldots , l_n}(\eps)}
\negthickspace \negthickspace \negthickspace \negthickspace
\negthickspace \negthickspace \negthickspace \negthickspace
\rho(x_n|x_{n-m}, \ldots, x_{n-1})\; dx_{n-m} \ldots dx_n \; ,
\end{align}
that a state inside some $\eps$-subset
of the state space is mapped onto some $\eps$-interval 
corresponding to the future, are treated.
$B_{l_{n-m}, \ldots , l_n}(\eps)$ is a resolution-dependent 
(m+1)-dimensional box as an element of the partition of the 
underlying embedding space.
The result are approximations to the true conditional
probability density, which vary only on spatial scales
which are larger than $\eps$. 
E.g., a model obtained for relatively small
$\eps$ has the potential to represent very fine structures in the
state space, but it suffers from poor statistics. 
Since for larger $\eps$ 
statistics gets better, but only coarser structures are resolved,
the optimal Markov order (and later on the optimal perforated
model), as well as prediction errors which will be discussed in Sec.
\ref{sec:conseqforpred}, depend on the spatial 
resolution \nolinebreak $\eps$.

Eventually we want to make plausible 
that the criterion Eq.(\ref{eq:critusualmarkovapprox}) for 
$m_{\operatorname{opt}}$ derived from information
theory really yields the optimal order of the Markov model
describing the underlying dynamics.
The conditional probabilities of Eq.(\ref{eq:condprobs})
corresponding to a Markov model of order $m$ are
estimated from a finite dataset. 
Hence they are subject to statistical errors, 
which are the larger the larger $m$. 
Exactly the same statistical errors of
conditional probabilities would lead to statistical errors 
of the redundancy $\Delta R_m$,
if we defined all information theoretic quantities through Shannon
entropies ($q=1$), and they enter indirectly the 
statistical errors of quantities based on the Renyi-entropy of order $q=2$
through $\Delta w_m$ (cmp.~Eqs.\,(\ref{eq:approxrelation}) 
and (\ref{eq:realizcardininneighborhood})).  
Hence the statistical error of the redundancy is 
related to uncertainty of the corresponding Markov model. 
Since furthermore with increasing conditioning 
the minimization of the ignored memory $Q_m$ 
is in accordance with the minimization of the modelling error 
of the Markov model
the plausibility argument is complete.

\begin{figure}[h]
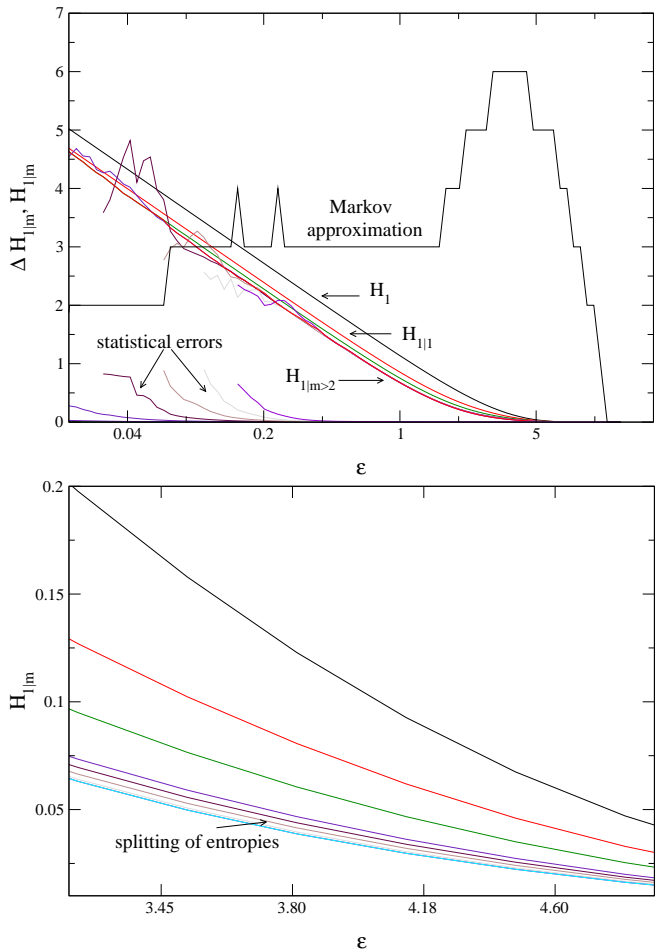

\bc
\includegraphics[width=8.6cm, angle=0]{./ar3_1_0k2_0k3_0k4.eps}
\includegraphics[width=8.6cm, angle=0]{./ar3_1_0k2_0k3_0k4zoom.eps}
\ec
\caption[]{\small\label{fig:ar3_usualmarkovapprox} (Color online)
Upper panel: Suggestion for resolution-dependent usual Markov approximations 
for the autoregressive process. 
Furthermore the conditional entropies $H_{1|m}(\eps)$
with varying conditioning as a function of the resolution and the 
corresponding resolution-dependent statistical error in the 
estimation of the conditional entropies are shown.
Lower panel: Zoom of the upper panel 
with the intention to make visible the splitting
of entropies for rather coarse resolutions, 
which is detected by the algorithm.}
\end{figure}

As an example the autoregressive (AR) process 
\beq
\label{eq:arpprocess}
x_{n+1}=\sum_{i=0}^{p-1}a_ix_{n-i}+\xi_{n+1} 
\eeq
of order $p=3$ with parameters 
$a_0=0.2$; 
$a_1=0.3$; 
$a_2=0.4$
is treated, i.e., a memory depth of three time steps is used. 
As usual $\xi_{n+1}$ is Gaussian white noise with unit variance and zero mean.
A dataset of length $N=50000$ is used.
The result is shown in Fig.\ref{fig:ar3_usualmarkovapprox}.

For intermediate resolutions the memory depth $m=p=3$
is exactly found with the algorithm, visible in the upper panel
of Fig.\ref{fig:ar3_usualmarkovapprox}.
For higher resolutions, i.e., smaller $\eps$, the statistical error 
dominates the criterion and a truncation for shorter 
Markov order is enforced.
This means that although the data stem from a process of Markov order 
$p=3$, for the given dataset size and chosen $\eps$ an $m=2$
model is superior when estimated from the data. 
For large $\eps$ a suggestion for a larger Markov order can be found.
The order of the Markov property given by Eq.(\ref{eq:arpprocess})
is also not preserved under coarse graining, which can be observed 
by the splitting of entropies with higher conditioning
in the lower panel of Fig.\ref{fig:ar3_usualmarkovapprox}, 
because for coarser resolutions 
the mapping onto discrete states becomes noticable and 
causes extra dependences 
among the involved random variables shifting 
information about the future into the further past. 
Since for coarse resolution the statistical
error is extremely small, the splitting of entropies is detected 
by the criterion as shown in the upper panel of 
Fig.\ref{fig:ar3_usualmarkovapprox}.

Whereas the application of the criterion given by 
Eq.(\ref{eq:critusualmarkovapprox}) was successful 
in the previous example, problems do arise in case of more 
general dynamics.
E.g., the discretized Mackey-Glass dynamics 
to be discussed in Sec.\ref{sec:mackey} leads to a memory
structure with omissions, i.e., certain intermediate time steps in
the past do not contribute to uncertainty reduction of the future.
Under the conditions of this section minimization of the modelling error 
is in general accompanied by large statistical errors
such that true joint minima of both types of errors are not 
accessed.
Hence a more subtle procedure for obtaining optimal 
Markov approximations should be necessary, in which the minimization
of the modelling error by contributions from the further past 
is not statistically suppressed. A notation for joint entropies 
on time series segments with omissions has to be introduced.
We call such situations 'perforated', which are worked out in the 
next section.
As we will see, on the other hand, the perforated framework introduces 
the new problem that with respect to a criterion for an optimal 
Markov approximation a  monotonicity of the relevant entropic 
quantities in a parameter 
as the Markov order $m$ in the former case  
describing all possible conditionings 
is not anymore available. A solution with a qualitatively slightly 
different generalized criterion, 
which nevertheless follows 
essentially the same idea as in this section, will be offered in 
Sec.\ref{sec:optgeneralizmarkovapprox}.

\section{Perforation}
\label{sec:perforatedness}
Whereas in Eq.(\ref{eq:renyiblockentropy}) the uncertainty 
of $m$ random variables corresponding to successive time steps in 
a time series is assessed, in this section a notational framework
for evaluation of uncertainties of random variables 
corresponding to arbitrary sets of time steps is introduced. 
Instead of the number $m$ of successive time steps, which is not 
anymore enough for characterization of the uncertainty-assessed set
of time steps, 
the relevant set has to be given explicitly. 
We will denote such sets of integers by ${\bf{J}}$ (or ${\bf{K}}$) 
and they will be called {\emph{perforated}},
if omissions of time steps are involved. 
E.g., Eq.(\ref{eq:renyizweientropy})
for the estimation of order-2-Renyi entropies has to be generalized 
for the perforated case by
\beq
\label{eq:estimperfentrop}
H_{{\bf{J}}}^{(2)}(N, \eps) = - \ln C_{{\bf{J}}}^{(2)}(N,\eps)\; ,
\eeq
where the vectors $\bx_i$, $\bx_k$
in Eq.(\ref{eq:correlationsum}) for the correlation sum 
adopt the perforation structure given by ${\bf{J}}$,
i.e., if ${\bf{J}}=\{j_1, j_2, \ldots , j_{|{\mathbf{J}}|}\}$, then 
the corresponding generalized delay vector with index $i$ 
reads $\bx_i=(x_{i+j_1}, x_{i+j_2}, \ldots , x_{i+j_{|{\mathbf{J}}|}})$.
This in general leads to non-standard embeddings.

Conditional entropies can be defined in general as
\beq
\label{eq:condentperf}
H_{{\bf{K}}|{\bf{J}}}^{(2)}(\eps) := H_{{\bf{K}}\cup {\bf{J}}}^{(2)}(\eps) 
- H_{{\bf{J}}}^{(2)}(\eps) \; .
\eeq
where ${\bf{K}}$ is a set of integers which is disjoint from ${\bf{J}}$.
This quantity in principle allows for the 
evaluation of entropies with noncausal conditioning.
In prediction situations the convention is made that the presence 
is indicated by the index zero. Hence a set ${\bf{J}}$ of conditioning indices 
in the past only consists of negative integers
$({\bf{J}} \subset \mathbb{Z}_0^-)$, 
which indicate the respective distances to the presence. 
With respect to optimal Markov approximations we are interested in 
single element sets ${\bf{K}}$. In this case the single element 
denoted by $f$ corresponds to a certain future time step, 
and Eq.(\ref{eq:condentperf}) reduces to
\beq
H_{\{f\}|{\bf{J}}}^{(2)}(\eps) := H_{\{f\}\cup {\bf{J}}}^{(2)}(\eps) 
- H_{{\bf{J}}}^{(2)}(\eps)
\; .
\eeq
With conditioning on full past for a single time step $f$ in the future 
the condtional entropy becomes $H_{\{f\}|\mathbb{Z}_0^-}(\eps)$.
As a special case of one step ahead the nonperforated conditional entropy 
with infinite conditioning of 
Eq.(\ref{eq:infiniteconditcondent}) is obtained:
\beq
H_{\{1\}|\mathbb{Z}_0^-}(\eps) \equiv H_{1|\infty} (\eps)\; .
\eeq
Under perforated circumstances
the ignored memory of Eq.(\ref{eq:ignoredmemory}) is redefined by
\beq
\label{eq:ignoredmemoryperf}
Q_{\{f\};{\bf{J}}}(\eps) :=
H_{\{f\}|{\bf{J}}}(\eps)-H_{\{f\}|\mathbb{Z}_0^-}(\eps) \; ,
\eeq
and the redundancy of Eq.(\ref{eq:redundancy}) now is obtained from
\beq
\label{eq:redundperf}
R_{\{f\};{\bf{J}}}(\eps):=H_1(\eps)-H_{\{f\}|{\bf{J}}}(\eps)\; .
\eeq
As a generalization of Eq.(\ref{eq:statisterrofnonperfredund}),
the statistical error of the redundancy in Eq.(\ref{eq:redundperf}), 
which will be essential for 
the novel criterion for optimal perforated Markov approximations,
still obtained from usual error propagation, reads
\begin{align}
\label{eq:perfredundancyerror}
& \Delta  R_{\{f\};{\mathbf{J}}}(N, \eps) \nonumber \\
& \! = \! \sqrt{[\Delta \! H_1(N, \eps)]^2
\!\! + \! [\Delta \! H_{\{f\}\cup {\bf{J}}}(N, \eps)]^2
\!\! + \! [\Delta \! H_{\bf{J}}(N, \eps)]^2 }
\end{align}
After having fixed the notational framework for 
a perforated treatment, a suitable generalization of the criterion 
for optimal usual Markov approximations of 
Sec.\ref{sec:optimusualmarkovapprox} can be given
such that simultaneous minimization of as well the modelling 
error as also the statistical error makes sense also for generalized 
dynamics containing inhomogeneously distributed memory in the past.
Moreover, the new notation in principle allows for a treatment of 
variable future time steps, jointly conditioned joint entropies, 
arbitrary omissions in conditionings, noncausal conditionings and
downsampling in a unified framework.

\section{A novel criterion for optimal generalized Markov approximations}
\label{sec:optgeneralizmarkovapprox}
Also in the perforated case we consider the two types of errors 
already discussed in the context of usual Markov approximations, 
i.e., the modelling error and the statistical error,
which have to be minimized jointly. 
As in Sec.\ref{sec:optimusualmarkovapprox}
the two errors are again quantified by the ignored memory, i.e., 
ignored potentially usable information, and the 
statistical error of redundancy, however, in this case 
with usage of variants of those quantities respecting perforatedness
as introduced in Sec.\ref{sec:perforatedness}. The minimization of the
single errors is again complementary in the number of conditioning indices,
but more subtle here, because in particular the ignored memory is not
only a function of the cardinality of the conditioning set, but 
depends explicitly on its single elements.

For finding the resolution-dependent optimal perforated Markov 
approximation, i.e., the optimal conditioning sets ${\bf{J}}^*(\eps)$, 
as the central criterion and most important formula of this paper 
it is demanded 
\beq
\label{eq:optimperfcriterion}
\qquad Q_{\{f\};{\bf{J}}}(\eps)+b \cdot \Delta R_{\{f\};{\bf{J}}}(N, \eps)
\stackrel{!}{=}{\min} \; , \qquad 
\eeq
where for given $\eps$ the minimum is taken in principle over all possible,
practically over all numerically accessible 
conditionings ${\bf{J}} \subset \mathbb{Z}_0^- $, instead of over all 
Markov orders $m$ as in Sec.\ref{sec:optimusualmarkovapprox}. 
The ignored memory $Q_{\{f\};{\bf{J}}}(\eps)$ in the perforated case was 
defined in Eq.(\ref{eq:ignoredmemoryperf})
and the statistical error of the redundancy 
$\Delta R_{\{f\};{\bf{J}}}(N, \eps)$ 
is obtained from Eq.(\ref{eq:perfredundancyerror}).
The parameter $b$ accounts for the weight 
of the statistical error of the redundancy in the criterion.
However, all results of Sec.\ref{sec:examplesperfmarkovapprox}
will be based on the choice $b=1$. 
A short discussion on balance factors $b \neq 1$
can be found in Sec.7 of \cite{holstdiss07}.
If the solution for a certain $\eps$ is not unique, 
it is taken in a second step the set ${\bf{J}}(\eps)$ as
${\bf{J}}^*(\eps)$ with 
\beq
\label{eq:optimperfcritsecondpart}
\min ({\bf{J}}(\eps)) = \max
\eeq
among the preselected ones.

The result is a resolution-dependent suggestion for 
optimal perforated Markov approximations.
The chosen criterion will obtain its justification by the 
ability to recover known models behind sufficiently large data sets
in a suitable intermediate interval of resolutions shown in 
Sec.\ref{sec:examplesperfmarkovapprox}. 

For the criterion of Eq.(\ref{eq:optimperfcriterion}) 
\begin{align}
\label{eq:perfcritwithsingleterms}
& Q_{\{f\};{\bf{J}}}(\eps)+b \cdot \Delta R_{\{f\};{\bf{J}}}(N, \eps) 
\nonumber \\
& \qquad =H_{\{f\}|{\bf{J}}}(\eps)-H_{\{f\}|\mathbb{Z}_0^-}(\eps) \nonumber \\
& \qquad \qquad \qquad +b\cdot
\sqrt{\Delta H_1^2(N, \eps) + \Delta H_{\{f\}|{\bf{J}}}^2(N, \eps)} 
\nonumber \\
& \qquad \stackrel{!}{=}\min
\end{align}
a simplified approximative representation can be given by
\beq
\label{eq:simplifiedcriterion}
H_{\{f\}|{\bf{J}}}(\eps)+b \cdot 
\Delta H_{\{f\}|{\bf{J}}}(N, \eps)\stackrel{!}{=}\mbox{min} \; ,
\eeq
because $H_{\{f\}|\mathbb{Z}_0^-}(\eps)$ and $\Delta H_1(N,\eps)$ 
are independent of $\bf{J}$ and hence act as constants for given 
resolution $\eps$.
Eq.(\ref{eq:simplifiedcriterion})
is a very good approximation of  
Eq.(\ref{eq:perfcritwithsingleterms}), 
since $\Delta H_1(N, \eps)$ is in general small compared to 
$\Delta H_{\{f\}|{\bf{J}}}(N, \eps)$.
The interpretation of this approximation of the criterion is that
the value of the conditional entropy including its statistical error has 
to be minimal.

\section{Examples}
\label{sec:examplesperfmarkovapprox}
In order to evaluate the ability of the introduced criterion for 
determination of optimal perforated Markov approximations, 
it is tested on data sets,
for which the structure of dependence is known.
The test is carried out with linear stochastic and with 
nonlinear deterministic dynamics. In the context of 
the example of the autoregressive process
furthermore the dependence of the output of the criterion 
on the length of the underlying dataset is explicitly addressed.

\subsection{Autoregressive processes}
\label{sec:findback_ar}
\subsubsection{Suggestion of the optimal perforated Markov approximation and 
comparison with the memory structure underlying the dataset}
\begin{figure}[h]
\bc
\includegraphics[height=8.6cm, angle=-90]{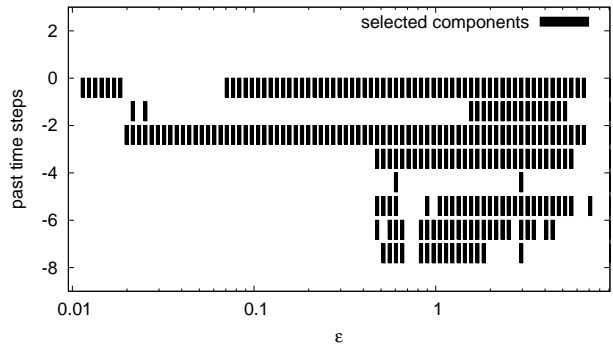}
\ec
\caption[]{\small
Resolution-dependent optimal perforated Markov model
for a dataset of $N=40000$ data points of an AR(3) process with $a_0=a_2=0.4$. 
The optimal conditioning structures can be found in vertical direction.
}
\label{fig:ar3_0k4_0_0k4_N40000_maxpsts7_fb} 
\end{figure}

The map of AR processes was given in Eq.(\ref{eq:arpprocess}).
For the first analysis a dataset of $N=40000$ data points 
is generated for a simple autoregressive 
process with parameters $a_0=a_2=0.4$, which 
fixes the structure of dependence in the iteration procedure. 
Parameters not mentioned are understood to be zero.
The time step with index '1' depends on the time steps given by 
the set ${\bf{J}}_0=\{-2, 0\}$.
A full search for the $\eps$-dependent optimal conditioning structure 
according to the criterion stated in Eq.(\ref{eq:optimperfcriterion}) is 
carried out, where additionally in case of the estimational result
$H_{\{f\}|{\bf{J}}}(\eps)<H_{\{f\}|\mathbb{Z}_0^-}(\eps)$
as a consequence of statistical fluctuations, what is 
theoretically impossible,
the estimated value of $H_{\{f\}|{\bf{J}}}(\eps)$ was replaced by the 
value of $H_{\{f\}|\mathbb{Z}_0^-}(\eps)$,
thus from Eq.(\ref{eq:ignoredmemoryperf})
avoiding negative $Q_{\{f\};{\bf{J}}}(\eps)$
in Eq.(\ref{eq:optimperfcriterion}) .
The result is shown in Fig.\ref{fig:ar3_0k4_0_0k4_N40000_maxpsts7_fb}.

A first result is that the found optimal conditioning structure 
${\bf{J}}^*(\eps)$ is indeed resolution-dependent.
Interpreting Fig.\ref{fig:ar3_0k4_0_0k4_N40000_maxpsts7_fb}, 
it is possible to extract three regimes: 

For high resolution, i.e. small $\eps$, the statistical errors 
of the entropy estimations are rather large, because in particular 
for longer conditioning fewer neighboring delay vectors for the 
estimation of the correlation sum can be found,
and hence the criterion is dominated 
by the statistical error of the redundancy, which causes 
perforated structures with fewer elements to be detected as optimal.
Interestingly, the single conditioning on the further past is selected 
as superior to single conditioning on the presence.
The statistical errors
of $H_{\{1\}|\{0\}}$ and of $H_{\{1\}|\{-2\}}$ are about the same, 
but the conditional entropy is estimated slightly smaller in the latter case.

For intermediate resolutions, the most interesting part of the plot,
the model behind the dataset is found, i.e.,
\beq
{\bf{J}}^*(\eps)={\bf{J}}_0\; ,
\eeq
because the statistical error is sufficiently small and the resolution
is sufficiently large that the information term dominates the
criterion without being disturbed by either statistical or
resolution effects.
Nevertheless, in this domain of resolutions the statistical error 
of the redundancy has the task to exclude all conditionings longer 
than necessary among those which are equal and optimal from the 
informational point of view.

For even coarser resolution, there is the domain of coarse graining 
splittings. 
The statistical error typically does not play a role anymore and the 
information term becomes influenced by resolution effects.
Even though the components 
do not carry information from the dynamical law, analyzing the dataset with
coarse resolution they are frequently chosen to appear in the optimal
perforated Markov model.
This is the same effect as shown already for the usual Markov approximation
of the autoregressive process in Fig.\ref{fig:ar3_usualmarkovapprox}.

The criterion is tested by the question if the components of 
conditioning ${\bf{J}}_0$ in the dynamics behind the generated dataset 
can be retrieved. In the rather simple case of autoregressive 
processes hence the criterion can be applied successfully.

An alternative and widely used approach to Gaussian time series data
is to directly fit the parameters of a linear model Eq.(20) to them.
Such routines minimize the variance of the residuals without
making use of information theoretic concepts.
For pre-selected model order $p$, using the whole dataset each single 
parameter $a_i$ is estimated explicitly, instead of only selecting 
components.
Hence, such fitting methods appear to be superior to what we propose 
here, and indeed for data from AR-processes, they are superior.
However, firstly, our goal is to identify the
relevant components of a delay vector, which includes the determination 
of the model order $p$, without pre-selection. 
Secondly, our approach is neither restricted to
data from linear models nor to Gaussian data, but develops its full
strength for nonlinear (stochastic) systems, as we will demonstrate
later.

\subsubsection{Dataset length dependence of the optimal perforated Markov model}
\begin{figure}[h]
\bc
\includegraphics[height=8cm, angle=-90]{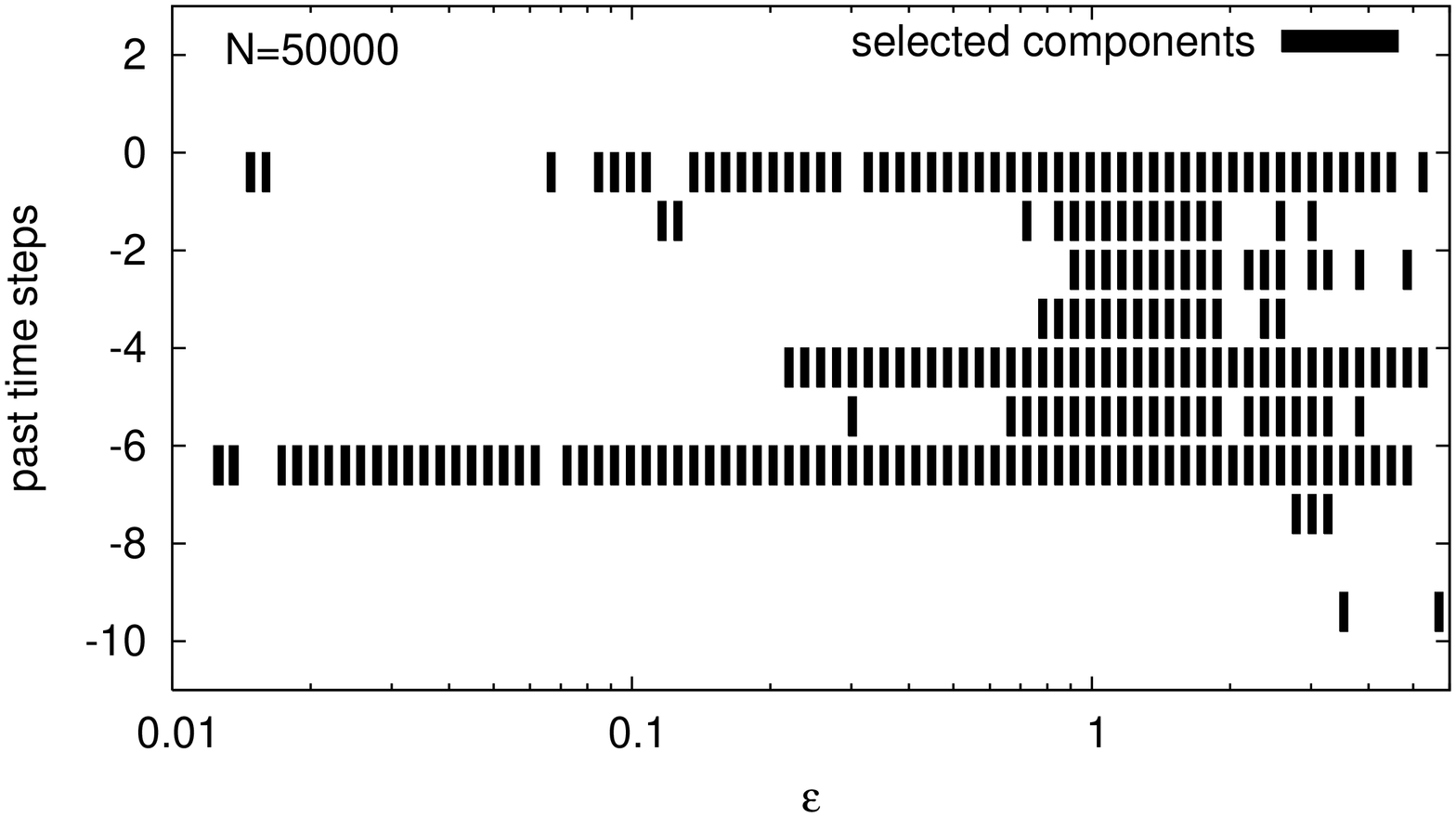} \\
\includegraphics[height=8cm, angle=-90]{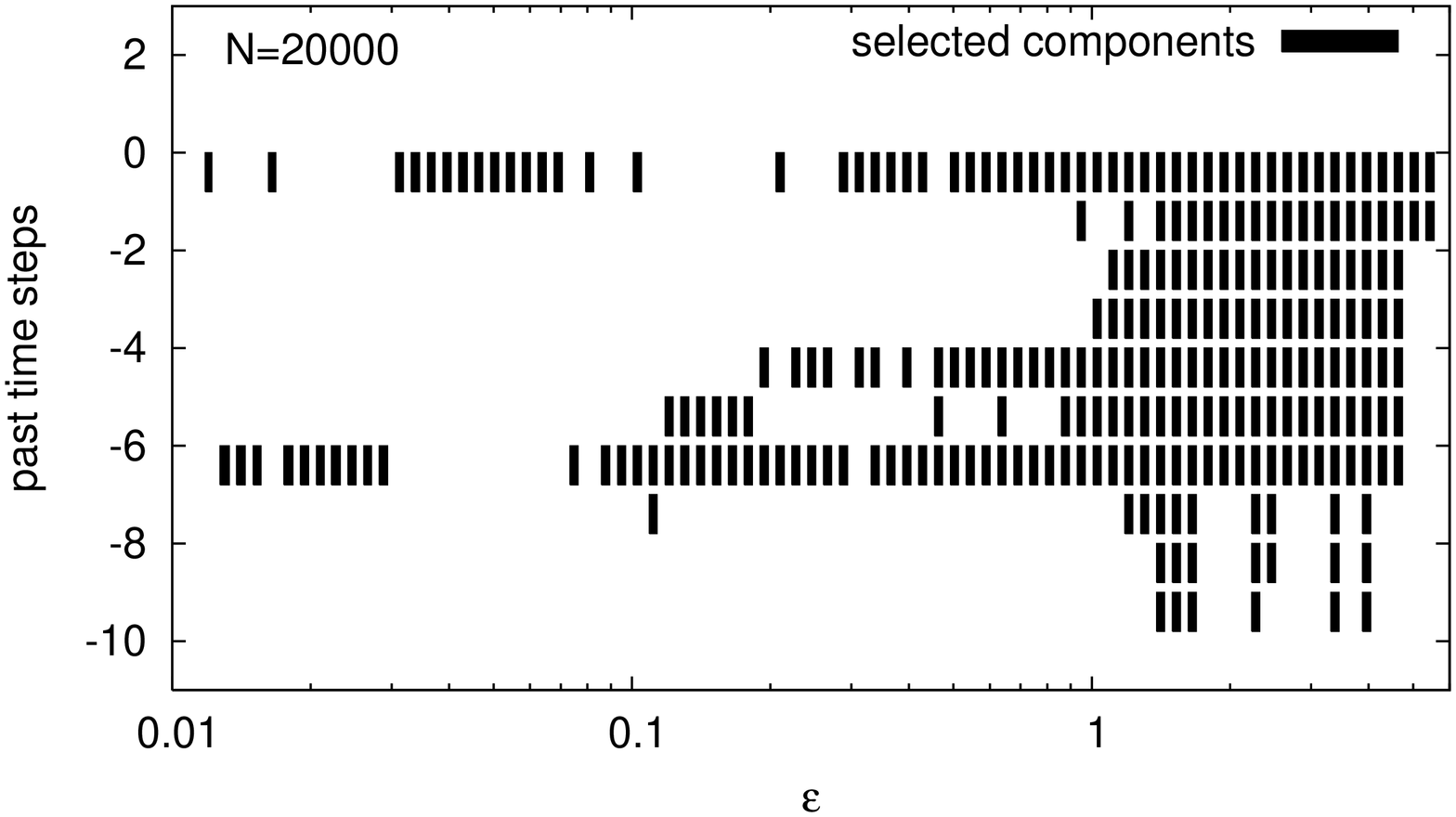} \\
\includegraphics[height=8cm, angle=-90]{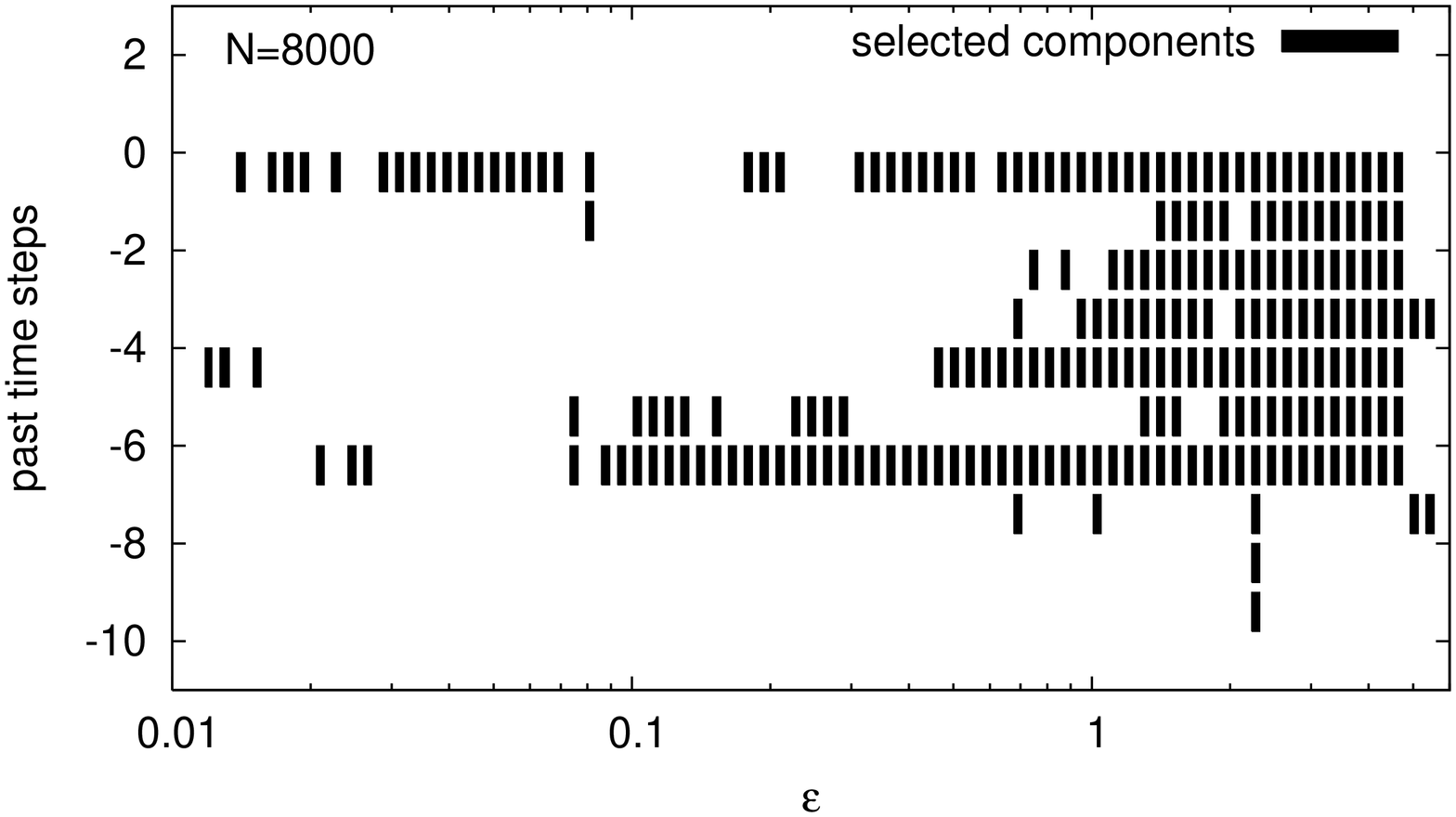} \\
\includegraphics[height=8cm, angle=-90]{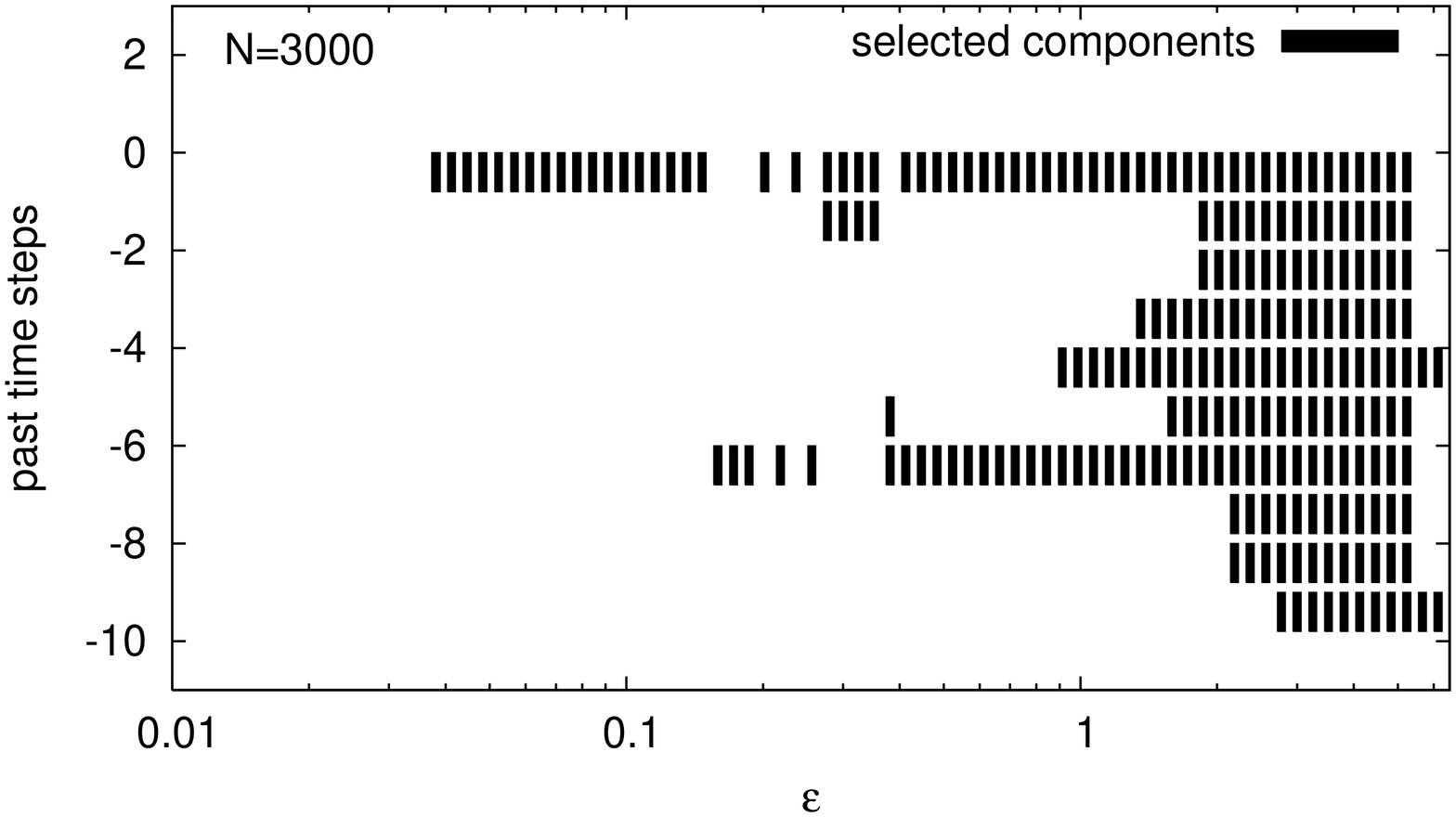}
\ec
\caption[]{\small\label{fig:ar7_046_changeddatasetlength}
Resolution-dependent optimal perforated Markov models of an AR(7) 
with coefficients $a_0=a_4=a_6=0.3$ under changed length $N$ of data sets.
The results are obtained from a full loop over all possible conditionings 
restricted only by the maximal number of 10 past time steps.
}
\end{figure}

Since an essential part of the criterion (\ref{eq:optimperfcriterion}) 
consists of a statistical error it is immediately clear that the result 
always depends crucially on the length of the underlying dataset.
The consequences of the influence of the length of the dataset 
are outlined in the following.
As the analyzed example a special AR(7) process 
\beq
x_{n+1}=0.3 \, x_n+0.3 \, x_{n-4}+0.3 \, x_{n-6} +\xi_n
\eeq
is used, for which the structure of dependence in the iteration procedure
can be described by the set ${\bf{J}}_0=\{-6, -4,0\}$.
With this iteration procedure data sets of different lengthes 
(N=3000, 8000, 20000, 50000) are generated and then analyzed 
with respect to the optimal resolution-dependent 
perforated Markov approximations.
The results are shown in Fig.\ref{fig:ar7_046_changeddatasetlength}.
It is possible to conclude that in the case of longer data sets
the time steps of memory in the used dynamics
can be retrieved on a broader interval of resolutions with higher 
reliability.
For shorter data sets the influence of the statistical error
in the criterion (\ref{eq:optimperfcriterion})
increases and the domain of dominance of the information term
is shifted to coarser resolutions seen in the selection of fewer 
components for the optimal model, where the statistical error term 
is dominant.
The structure of conditioning ${\bf{J}}_0$ of the
underlying dynamics becomes blurred, if the domain
of dominance of the statistical error starts to touch 
the domain of coarse graining effects for sufficiently short data sets. 
In the bottom right panel of 
Fig.\ref{fig:ar7_046_changeddatasetlength} this case is almost reached.

In the hypothetical case of infinite dataset length all statistical 
errors become zero for all resolutions and the criterion 
(\ref{eq:optimperfcriterion}) is governed by the ignored memory.
Optimality is selected for minimal modelling error quantified by 
vanishing ignored memory.
If memory ranges infinitely far into the past, then a Markov approximation
is always accompanied by a loss of information.
According to the criterion a Markov approximation of finite order 
can thus not be selected as optimal.
If the range of memory is finite into the past, a Markov approximation 
is possible where no information is found in the further past, but 
it would not be necessary, 
because components of the past without information about the future
nevertheless kept do not diminish the quality of the model
with respect to the first part of the criterion 
in Eq.(\ref{eq:optimperfcriterion}) 
in case of infinite data sets.
The second part of the criterion given in 
Eq.(\ref{eq:optimperfcritsecondpart}) 
decides for the shortest conditioning in the set of degenerated 
selected perforated Markov models.

\subsection{Mackey-Glass dynamics}
\label{sec:mackey}
\begin{figure}[h]
\bc
\includegraphics[height=7.85cm, angle=-90]{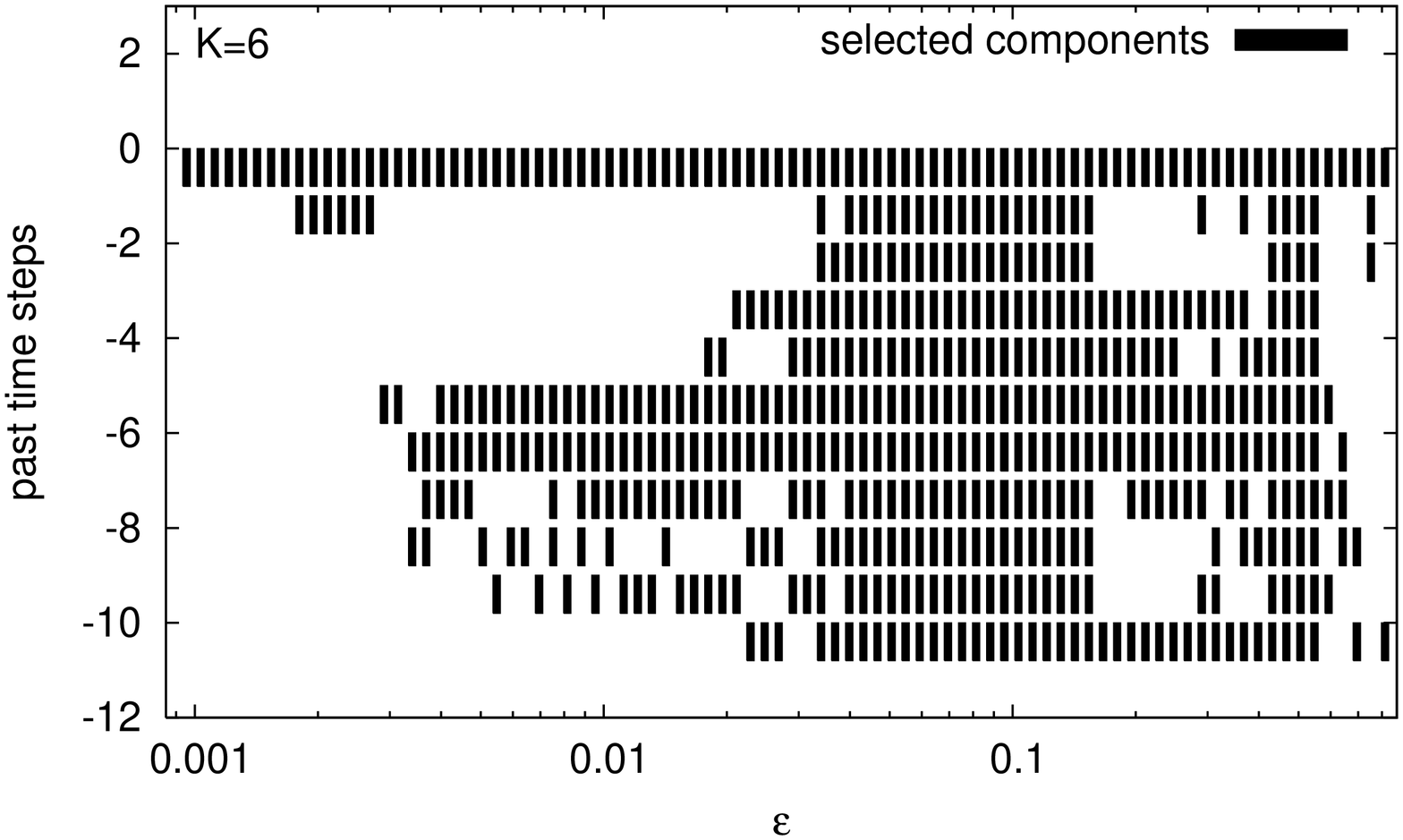} \\
\includegraphics[height=7.85cm, angle=-90]{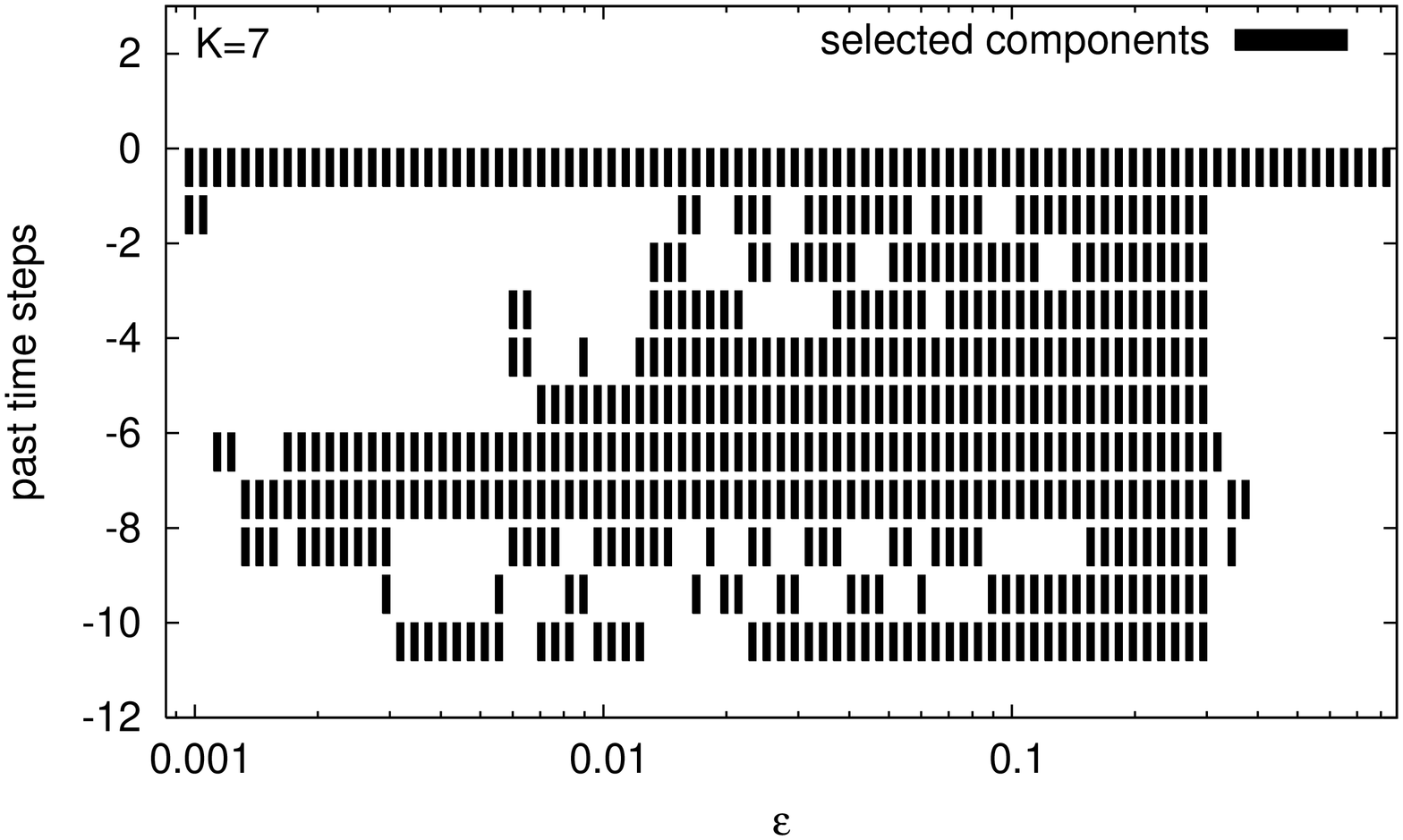} \\
\includegraphics[height=7.85cm, angle=-90]{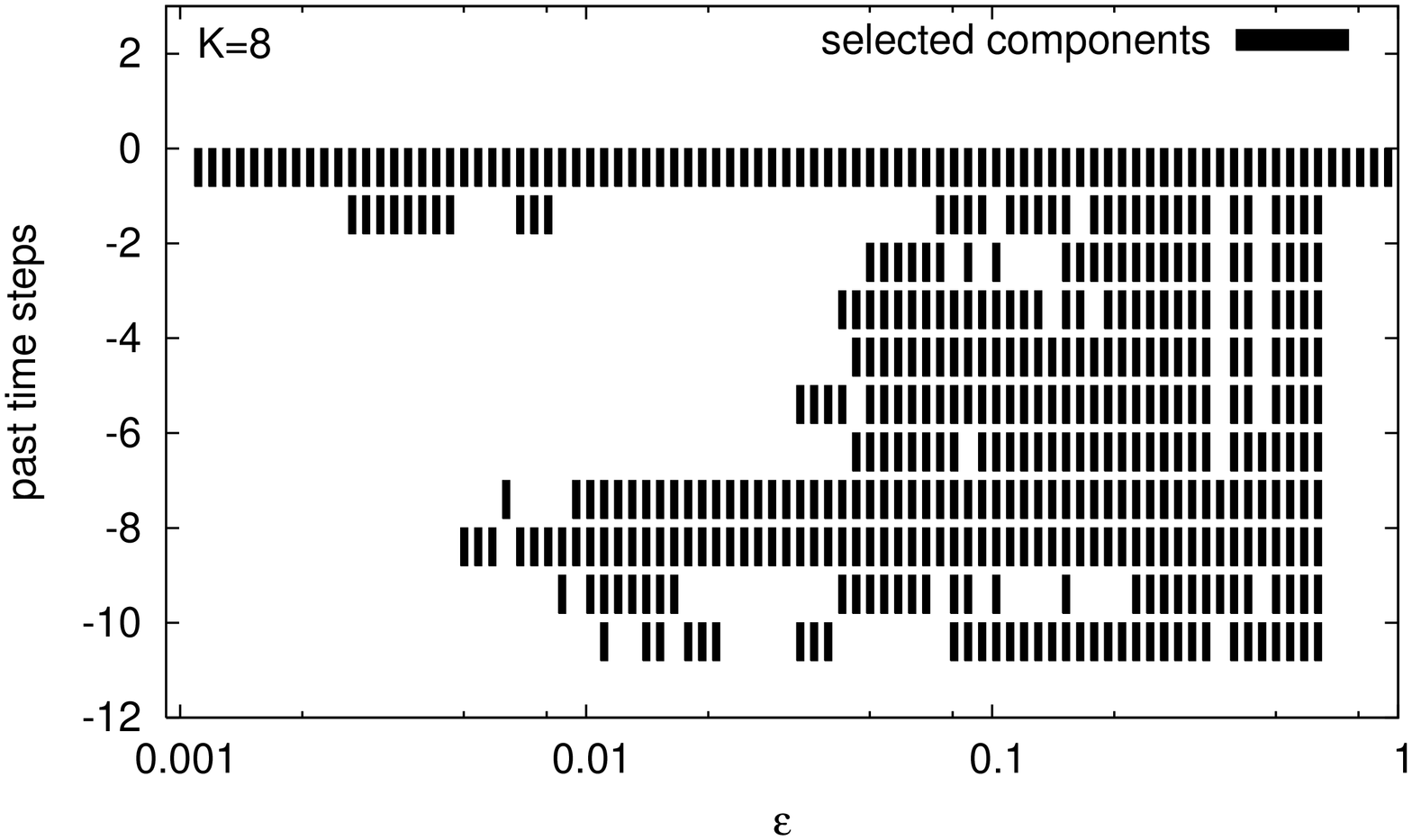} \\
\includegraphics[height=7.85cm, angle=-90]{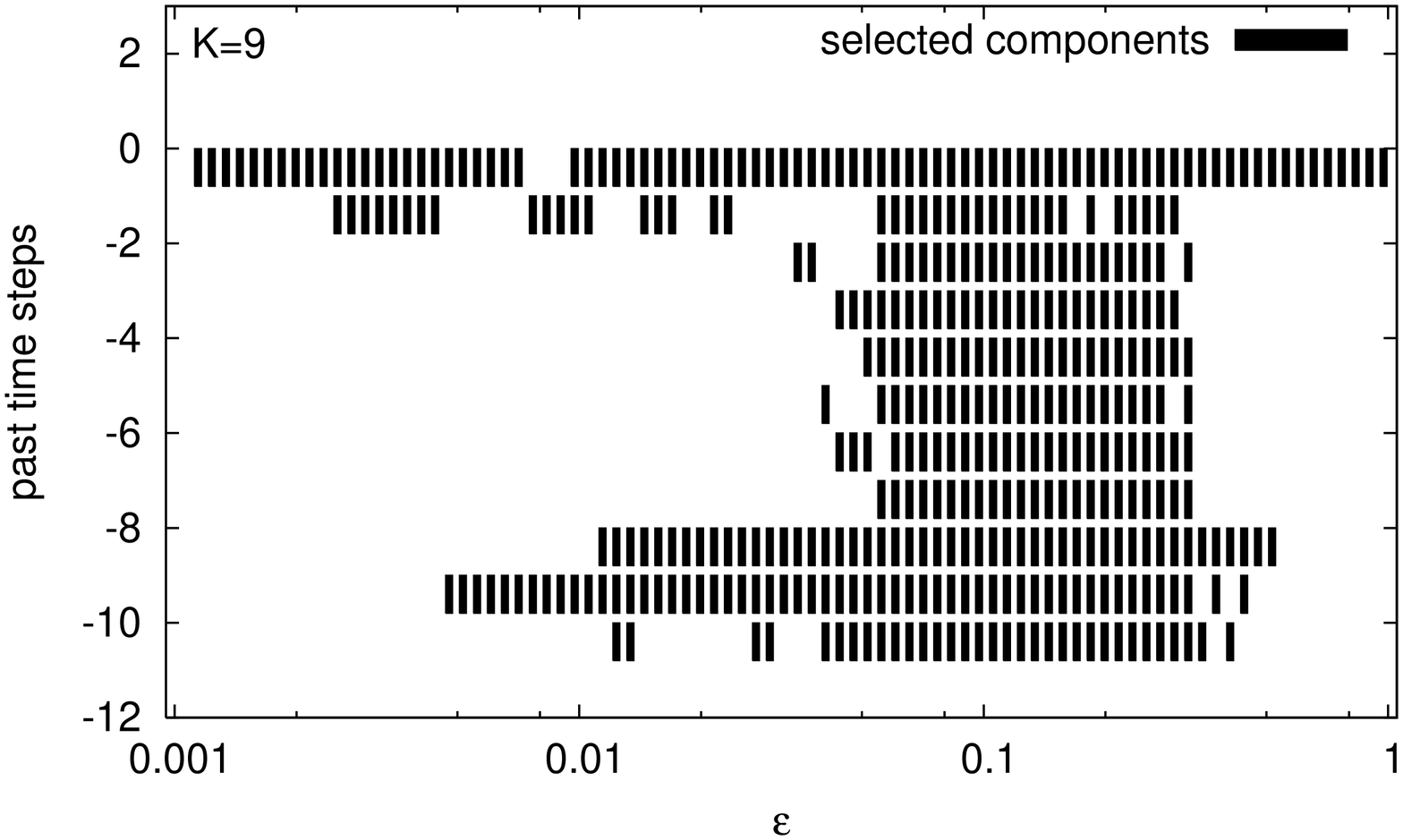}
\ec
\caption[]{\small\label{fig:mackeygl_optgdv_delvariabel_mod9}
Resolution-dependent optimal perforated Markov models for the 
Mackey-Glass dynamics with different effective delays $K$.}
\end{figure}

As a second example for testing the performance of the criterion 
(\ref{eq:optimperfcriterion}) we analyze the 
Mackey-Glass dynamics \cite{mackeyglass77}
given by
\beq
\dot{x}(t)=\frac{ax(t-\tau)}{1+[x(t-\tau)]^c}-bx(t) \; ,
\eeq
a time-continuous nonlinear deterministic example with memory.
The state at time $t$ depends explicitly on the state at time $t-\tau$.
Mackey-Glass dynamics is a representative of the class of 
delay differential equations, a subset
of the set of infinite-dimensional dynamical systems. It serves as a model 
for the regeneration of white blood cells for patients with leucemia.
Discretized, the equation of motion reads
\beq
\label{eq:discretizedmackey}
x_{n+1}=(1-b \Delta t) x_n + \frac{ax_{n-k}}{1+x_{n-k}^c} \Delta t
\eeq
with the delay
\beq
k=\frac{\tau}{\Delta t} \in \mathbb{N} \; .
\eeq
Typical parameter values \cite{grassproc83c, farmer82} are
\beq
a=0.2\, , \qquad b=0.1\, , \qquad c=10 \; .
\eeq
As an example taking $\Delta t=0.01$ time units, 
a delay of e.g. $k=1800$ time steps 
leads to a time delay of $\tau = 18$ time units. 
For $\tau > 16.8$ time units it is known that 
the dynamics is essentially chaotic. 
Using every 300th time step in the dataset to analyze
leads to an effective delay of $K=6$ time steps.
For the following analysis, data sets of 12000 effective 
data points are used.
\begin{figure}[h]
\bc
\includegraphics[width=8.6cm, angle=0]{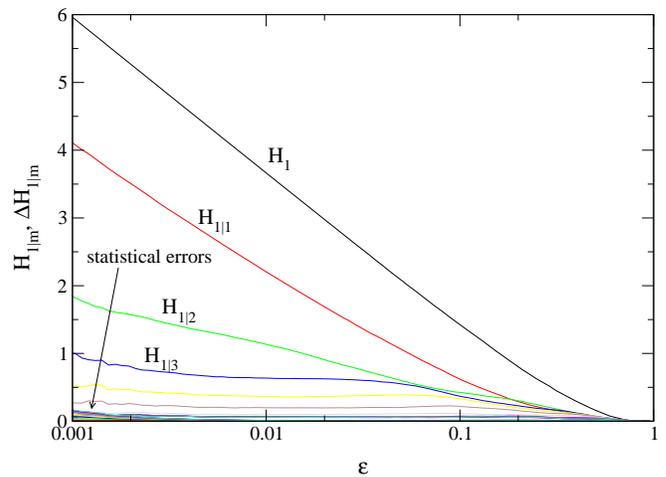}
\ec
\caption[]{\small\label{fig:mackeygl_entropies_del6} (Color online)
Conditional entropies of the Mackey-Glass dynamics with 
effective delay $K=6$}
\end{figure}
Even though in Fig.\ref{fig:mackeygl_entropies_del6}
for usual (nonperforated) conditional entropies 
the delay is invisible, the entropic-statistical criterion 
(\ref{eq:optimperfcriterion}) selects it.
This is seen in Fig.\ref{fig:mackeygl_optgdv_delvariabel_mod9},
where a whole series of optimal perforated Markov models
for different effective delays is shown.
The right part of the panels is again subject to coarse graining effects.
For higher resolution more structure is visible. 
The most important point to stress is that all panels have in common
that there is an interval of resolutions, where the optimal perforated Markov
model contains omissions behind the first step of conditioning and
the first following time step taken into account is exactly the time 
step corresponding to the effective delay of the dynamics.
The index 0 is always part of ${\bf{J}}^*(\eps)$, because of the 
$x_n$-term in Eq.(\ref{eq:discretizedmackey}).

Concluding, the very long range of the memory of the Mackey-Glass system
requires strong downsampling, from which the complication arises that
the resulting effective memory underlies some smearing effects.
Nevertheless, since it was detected by the criterion, also this example has
to be interpreted as a successful test of the criterion.

Without going into detail here it should be mentioned that in 
\cite{holstdiss07} various further variants for the selection 
of conditioning components as e.g. a restricted cardinality of 
conditioning components or a priori omissions of indices 
were suggested, in order to reach the further past for detection of 
potential memory.

\section{Consequences for prediction}
\label{sec:conseqforpred}
\subsection{Point prediction and prediction error}
\label{sec:pointpredandprederr}
General point prediction one time step into the future reads
\beq
\label{eq:pointprediction}
\hat{x}_{n+1} = F(\bx_n)
\eeq
with a suitably chosen function $F$.
The average quality of predictions can be evaluated by an accuracy measure.
We choose the root mean squared (rms) prediction error given by
\beq
\label{eq:predictionerror}
\hat{e}=\sqrt{\overline{(x_{n+1}-\hat{x}_{n+1})^2}} \; .
\eeq
As a consequence the {\em{mean value}} of the estimated distribution
of $X_{n+1}$, the random variable corresponding to the
measured value $x_{n+1}$, is the optimal $F$.
This distribution is estimated by a selected set of $x_{k+1}$,
which are obtained from those $\bx_k$, which are in some sense 
suitably related to $\bx_n$. 
A decision, what a 'suitable relation' should be, is not immediately given by
Eq.(\ref{eq:predictionerror}) and has to be made additionally.
Another possible accuracy measure could be the mean absolute error, which 
would lead to an optimal $F$ given by the median.
In general the prediction error depends on the lead time 
(time into the future), the dataset length $N$, the noise 
in the dynamical modeling $F$ and possibly on the resolution $\eps$.

\subsection{Locally constant prediction with generalized delay vectors}

A special point prediction used in the following, which is 
locally constant (cmp.~the zeroth order predictor in \cite{farmer87}) 
and perforated, reads
\begin{align}
\label{eq:pointpredictionlocalperf}
\hat{x}_{n+1} (\eps) &= \frac{\sum_{k\neq n}
\Theta (\eps- \| P_{\bf{J}}\bx_n-P_{\bf{J}}\bx_k\|)\cdot x_{k+1}}
{\sum_{k \neq n} 
\Theta (\eps-\|P_{\bf{J}}\bx_n-P_{\bf{J}}\bx_k\|)}
\nonumber \\
&= \! \frac{1}{|\{P_{\bf{J}}\bx_{k\neq n} \in
{\mathcal{U}}(\eps, P_{\bf{J}}\bx_n) \} |} \!\!
\sum_{P_{\bf{J}}\bx_{k \neq n} \in
{\mathcal{U}}(\eps, P_{\bf{J}}\bx_n)} \!\!\!\!\!\!\!\!
x_{k+1} \; .
\end{align}
$P_{\bf{J}}$ is the projection operator onto the perforation structure 
given by the set ${\bf{J}}$ already encountered in 
Sec.\ref{sec:perforatedness} and
${\mathcal{U}}(\eps, P_{\bf{J}}\bx_n)$ is the 
$\eps$-neighborhood of the vector $P_{\bf{J}}\bx_n$
introduced in Eq.(\ref{eq:epsneighborhood}).
Apart from the perforatedness the method is also called the 
Lorenz method of analogues: The predicted future value 
is the mean of the known futures of similar states from the past.
We will study explicitly the resolution dependence of the 
corresponding prediction error:
\beq
\label{eq:prederrorlocal}
\hat{e}(\eps)=\sqrt{\overline{(x_{n+1}-\hat{x}_{n+1}(\eps))^2}}\; .
\eeq

\subsection{Example: Prediction from optimal perforated Markov model for generalized Henon dynamics}
After having found the optimal resolution-dependent 
perforated Markov models from the criterion 
(\ref{eq:optimperfcriterion}) with the suitable balance $b$, the corresponding
component structures given by ${\bf{J}}^*(\eps)$
can be used for the calculation of 
point predictions according to Eq.(\ref{eq:pointpredictionlocalperf})
and rms prediction errors according to Eq.(\ref{eq:prederrorlocal}).
In the following the prediction error corresponding to the optimal 
perforated Markov model, i.e., conditioning in the sense of the optimal 
generalized delay vector (GDV), is compared to the minimum of the 
prediction error of usual standard embeddings (1 - 5 delay vector (DV) 
components in presence and past; delay of 1 - 5 time steps).
As an example we treat the generalized H\'{e}non map 
\beq
\label{eq:generalizhenon}
y_{n+1}=a-y_{n-K+2}^2-cy_{n-K+1} \; ,
\eeq
a simple chaotic system introduced 
by Baier\&Klein 
in \cite{baier90}.
In general it contains longer memory than the usual H\'{e}non map
\beq
\label{eq:usualhenon}
x_{n+1}=1-\alpha x_n^2+\beta x_{n-1} \; ,
\eeq
which is obtained from the generalized 
H\'{e}non map in the case of $K=2$ from the transformation $y=ax$,
$a=\alpha$ and $c=-\beta$.
The nonlinearity still arises from one single quadratic term.
The coefficients are chosen to be
$a=1.76$ and $c=0.1$.
From comparison of the coefficients $1$ vs. $c$ of the non-constant terms 
of Eq.(\ref{eq:generalizhenon})
it is possible to see in this case that the linear term is suppressed 
in importance.
From the choice of the delay $K=4$ the structure of dependence can be 
indicated by the set ${\bf{J}}_0=\{-3,-2\}$.

\begin{figure}[h]
\bc
\includegraphics[height=9cm,
angle=-90]{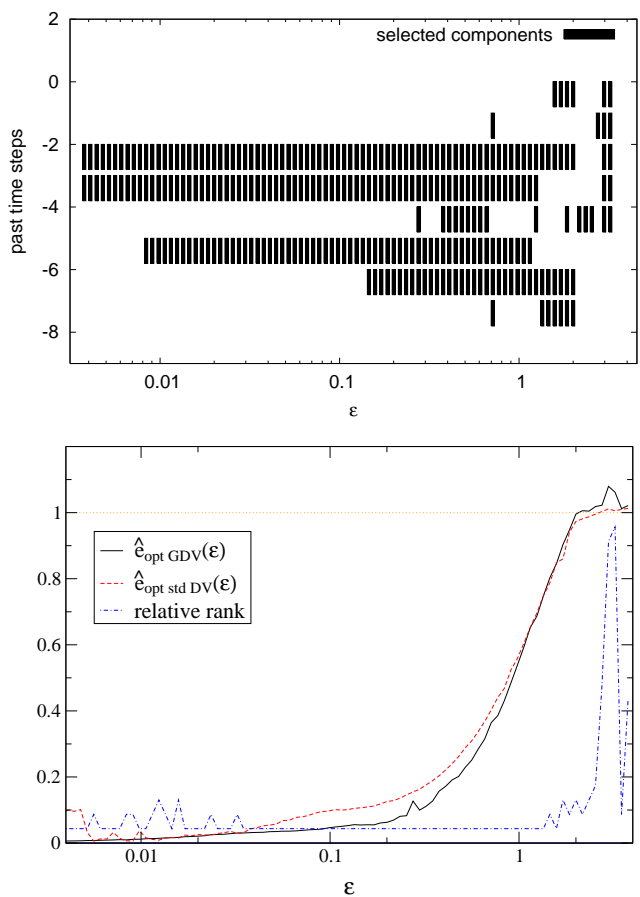}
\vglue 0.3cm
\hglue 0.34cm
\includegraphics[height=6.0cm,
angle=0]{./genhen_del4_a1k76_b0k1_10000_hinfty0_balance4_prederr.eps}
\ec
\caption[]{\small\label{fig:genhendel4_optgdv_prederr} (Color online)
Upper panel:
Optimal resolution-dependent perforated Markov model (${\bf{J}}^*(\eps)$)
for a dataset of $N=10000$ data points of the generalized H\'{e}non 
with delay $K=4$. 
Lower panel: Resolution-dependent prediction error 
$\hat{e}_{\operatorname{opt \, GDV}}(\eps)$
from 
${\bf{J}}^*(\eps)$,
minimal prediction error $\hat{e}_{\operatorname{opt \, std \, DV}}(\eps)$ 
of standard delay vectors and 
relative rank of the prediction error from 
${\bf{J}}^*(\eps)$
in the list of prediction errors from standard embeddings.
}
\end{figure}

In Fig.\ref{fig:genhendel4_optgdv_prederr} results for 
prediction (lower panel) 
from optimal perforated Markov models (upper panel) 
are shown for the generalized H\'{e}non map for 
a balance factor of $b=4$ in (\ref{eq:optimperfcriterion})
which favors models with fewer components.
It is seen that the prediction error from the optimal perforated 
Markov model is smaller than the minimum of prediction errors
from standard embeddings. This serves as a justification 
for the introduction of perforatedness into the framework of 
Markov approximations and for practical applicability of the 
criterion (\ref{eq:optimperfcriterion}) for prediction purposes.

\section{Conclusion}
\label{sec:conclusion}
For dynamics with potentially infinite memory, 
e.g. from projection of stochastic dynamics into one measurement quantity,  
novel criteria for optimal Markov approximations were introduced.
It was realized that essentially two types of errors are relevant:
First, a modelling error, quantified by the ignored memory, 
and second, a statistical error of uncertainty reduction, 
quantified by the statistical error of the redundancy.

Usually Markov approximations 
are accompanied by losses of information, which become 
less the more memory is taken into account. Exactly the opposite holds
for the statistical error of the uncertainty reduction, because 
a larger Markov order causes stronger restrictions in neighbor search 
algorithms responsible for larger statistical errors in the estimation 
of entropies and hence also of the redundancy.
The rather simple idea behind the criterion for usual Markov approximations 
is that it makes no sense to further reduce ignored memory if the statistical 
error of the uncertainty reduction is already larger.
Here the monotony properties of the involved quantities were used 
in the mathematical formulation of the criterion.

Even though this criterion was successfully applicable on simple dynamics,
problems arise from the huge statistical errors for high cardinality
of conditioning sets for dynamics with long range and inhomogeneously 
distributed memory. Hence, a generalization to a perforated case, where 
omissions of time steps in the past have to be allowed, was needed.
A generalized notational framework of information theory in time series 
analysis was developed, which in principle allows for a {\em{unified}}
description of variable future time steps ahead, jointly conditioned 
joint uncertainties, regular perforation (downsampling) and arbitrary 
irregular perforation with the tools of information theory. 
On this basis a novel criterion for optimal perforated Markov 
approximations was introduced, in which the selection algorithm for 
relevant conditioning components took into account the nonexistence 
of monotony properties of the modelling error in the cardinality of 
the conditioning set.
 
The perforated criterion was successfully tested for linear stochastic 
(AR) and nonlinear deterministic (Mackey-Glass) dynamics.
It was found that the optimal perforated Markov model is 
resolution-dependent. For certain intervals of intermediate resolution the 
memory structure of the dynamical law was retrieved
by the suggested criterion indicating the functional capability 
to yield suitable Markov approximations. For small resolutions
coarse graining effects are clearly seen and for fine resolutions 
from statistical reasons fewer conditioning components are selected.
The importance of the dependence on the length of the underlying dataset 
was pointed out.

Since the methods are based exclusively on quantities from 
information theory and statistical errors in their estimation,
in particular the perforated variant is applicable to a 
broad class of dynamics. 
This is especially useful 
for an analysis of data sets, where it is not allowed to assume 
nice properties like, e.g., linearity.
The explicit calculation of the statistical error of entropies made 
accessible those criteria based only on entropies and its derived 
quantities.
In spite of the success of the criterion on the example dynamics, 
it has to be mentioned that nonstationarity and intermittency 
still remain as problems.

For locally constant and perforated point prediction an explicitly 
resolution-dependent root mean squared prediction error was 
introduced. For certain resolutions an improvement of the 
rms prediction error from the resolution-dependent
optimal perforated Markov model in comparison with the 
rms prediction error from standard embeddings 
was seen in the example of the generalized H\'{e}non map.

%
%
%

\end{document}